\documentclass[12pt]{article}
\usepackage{amsmath,amssymb}
\usepackage{amscd}
\newcommand{\eqdef}{\stackrel{\text{def}}{=}}
\newcommand{\n}{\nonumber \\}
\newcommand{\bm}{\boldsymbol}
\newcommand{\ignore}[1]{}

\allowdisplaybreaks[4]

\begin{document}

\baselineskip=20pt

\newfont{\elevenmib}{cmmib10 scaled\magstep1}
\newcommand{\preprint}{
   \begin{flushleft}
     \elevenmib Yukawa\, Institute\, Kyoto\\
   \end{flushleft}\vspace{-1.3cm}
   \begin{flushright}\normalsize \sf
     DPSU-10-2\\
     YITP-10-19\\
   \end{flushright}}
\newcommand{\Title}[1]{{\baselineskip=26pt
   \begin{center} \Large \bf #1 \\ \ \\ \end{center}}}
\newcommand{\Author}{\begin{center}
   \large \bf S.~Odake${}^a$ and R.~Sasaki${}^b$ \end{center}}
\newcommand{\Address}{\begin{center}
     $^a$ Department of Physics, Shinshu University,\\
     Matsumoto 390-8621, Japan\\
     ${}^b$ Yukawa Institute for Theoretical Physics,\\
     Kyoto University, Kyoto 606-8502, Japan
   \end{center}}
\newcommand{\Accepted}[1]{\begin{center}
   {\large \sf #1}\\ \vspace{1mm}{\small \sf Accepted for Publication}
   \end{center}}

\preprint
\thispagestyle{empty}
\bigskip\bigskip\bigskip

\Title{Exceptional Askey-Wilson type polynomials through Darboux-Crum
transformations}
\Author

\Address

\begin{abstract}
An alternative derivation is presented of the infinitely many exceptional
Wilson and Askey-Wilson polynomials, which were introduced by the present
authors in 2009. Darboux-Crum transformations intertwining the discrete
quantum mechanical systems of the original and the exceptional polynomials
play an important role. Infinitely many continuous Hahn polynomials are
derived in the same manner.
The present method provides a simple proof of the shape invariance of
these systems as in the corresponding cases of the exceptional Laguerre
and Jacobi polynomials.
\end{abstract}

\section{Introduction}
\setcounter{equation}{0}

In a previous paper \cite{os17}, we have derived {\em infinitely many
exceptional orthogonal polynomials\/} related to the Wilson and
Askey-Wilson polynomials \cite{askey,ismail,koeswart} as the solutions
of infinitely many shape invariant \cite{genden} thus exactly solvable
{\em discrete\/} quantum mechanics in one dimension
\cite{os4,os6,os7,os13,os12}. The corresponding Hamiltonians are the
deformations of those for the Wilson and Askey-Wilson polynomials
\cite{os4,os7,os13} in terms of a degree $\ell$ ($\ell=1,2,\ldots$)
eigenpolynomial with twisted parameters \cite{os17}.
In discrete quantum mechanics, the Schr\"odinger equation is a second
order {\em difference\/} equation, which reduces, in known cases of
exactly solvable examples, to the difference equation satisfied by the
polynomials belonging to the Askey scheme of hypergeometric orthogonal
polynomials and their $q$-analogues.
As stressed in our previous publications, various concepts and formulas
of these Askey scheme polynomials can be understood and formulated in
a unified fashion through the framework of quantum mechanics;
that is the eigenvalue problem of a hermitian (self-adjoint) linear
operator (the Hamiltonian) in a certain Hilbert space.
The Hamiltonian corresponding to a particular polynomial, say the Wilson
or the Askey-Wilson polynomial, is specified by a choice of the potential
function \eqref{Vform} as in the ordinary quantum mechanics.

In this paper we present an alternative derivation of the infinitely
many exceptional Wilson and Askey-Wilson polynomials in terms of
Darboux-Crum transformations \cite{darboux,crum} {\em intertwining\/}
the Hamiltonians of the original Wilson/Askey-Wilson polynomials with
that of the corresponding exceptional polynomials.
The same method is applied to the continuous Hahn polynomials
\cite{koeswart} to construct the new infinitely many exceptional
continuous Hahn polynomials indexed by a positive even integer $\ell$.
This is a discrete quantum mechanics version of the recent work \cite{stz}
which derives the four families of infinitely many exceptional
Laguerre/Jacobi polynomials \cite{os16,os19} in terms of Darboux-Crum
translations {\em intertwining\/} the Hamiltonians of the exceptional
polynomials with those of the well known exactly solvable Hamiltonians
of the radial oscillator and the Darboux-P\"oschl-Teller potentials
\cite{infhul,susyqm,dpt}.

The concept of exceptional ($X_{\ell}$) polynomials was introduced by
Gomez-Ullate et al \cite{gomez} in 2008 as a new type of orthogonal
polynomials satisfying second order differential equations.
The $X_{\ell}$ polynomials start with degree $\ell$ ($\ell=1,2,\ldots$)
instead of the degree zero constant term, thus avoiding the constraints
by Bochner's theorem \cite{bochner}.
They constructed the $X_1$ Laguerre and Jacobi polynomials
\cite{gomez}, which are the first members of the infinitely many
exceptional Laguerre and Jacobi polynomials introduced by the present
authors \cite{os16,os19} in 2009.
Quesne gave a quantum mechanical formulation based on shape invariance
\cite{quesne} and later she introduced another set of $X_2$ polynomials
\cite{quesne2}.

This paper is organised as follows. In section two we review the discrete
quantum mechanical systems of the continuous Hahn, Wilson and Askey-Wilson
polynomials in \S\,\ref{sec:org.sys} together with those of the
corresponding exceptional polynomials in \S\,\ref{sec:deformed.sys}.
The properties of the deforming polynomial $\xi_{\ell}$ are discussed in
some detail in \S\,\ref{sec:xil_prop}.
They are important for deriving various results in the subsequent section.
The main part of the paper, the Darboux-Crum transformations intertwining
the original and the deformed systems are discussed in section three.
The final section is for a summary and comments including the
annihilation/creation operators and Rodrigues type formulas for the
exceptional polynomials.

\section{The original and deformed systems}
\setcounter{equation}{0}

Here we first recapitulate the shape invariant, therefore solvable,
systems whose eigenfunctions are described by the orthogonal polynomials;
the continuous Hahn, Wilson and Askey-Wilson polynomials \cite{koeswart},
to be abbreviated as cH, W and AW, respectively.
See \cite{os4} and \cite{os13} for the discrete quantum mechanical
treatment of these polynomials.
It is based on the factorisation method and generalisation of the
Darboux-Crum transformations \cite{darboux,crum}. See also the related
papers \cite{deift}. The factorisation of the Askey-Wilson polynomial
was also discussed in \cite{kalnins}.
Then the deformed systems corresponding to the exceptional Askey type
polynomial are summarised in \S\,\ref{sec:deformed.sys}.

\subsection{The original systems}
\label{sec:org.sys}

Here we summarise various properties of the original Hamiltonian systems
to be compared with the deformed systems which will be presented in
\S\,\ref{sec:deformed.sys}.
Let us start with the Hamiltonians, Schr\"{o}dinger equations and
eigenfunctions ($x_1<x<x_2$, $p=-i\frac{d}{dx}$):
\begin{align}
  &\mathcal{A}(\bm{\lambda})\eqdef
  i\bigl(e^{\frac{\gamma}{2}p}\sqrt{V^*(x;\bm{\lambda})}
  -e^{-\frac{\gamma}{2}p}\sqrt{V(x;\bm{\lambda})}\,\bigr),\n
  &\mathcal{A}(\bm{\lambda})^{\dagger}\eqdef
  -i\bigl(\sqrt{V(x;\bm{\lambda})}\,e^{\frac{\gamma}{2}p}
  -\sqrt{V^*(x;\bm{\lambda})}\,e^{-\frac{\gamma}{2}p}\bigr),\\
  &\mathcal{H}(\bm{\lambda})\eqdef
  \mathcal{A}(\bm{\lambda})^{\dagger}\mathcal{A}(\bm{\lambda}),
  \label{origham}\\
  &\mathcal{H}(\bm{\lambda})\phi_n(x;\bm{\lambda})
  =\mathcal{E}_n(\bm{\lambda})\phi_n(x;\bm{\lambda})\quad
  (n=0,1,2,\ldots),\\
  &\phi_n(x;\bm{\lambda})=\phi_0(x;\bm{\lambda})P_n(\eta(x);\bm{\lambda}).
\end{align}
The set of parameters $\bm{\lambda}=(\lambda_1,\lambda_2,\ldots)$ are
\begin{align}
  \text{cH}:\quad&\bm{\lambda}\eqdef(a_1,a_2),\quad
  \text{Re}\,{a_i}>0 \ \ (i=1,2),
  \label{paracH}\\
  \text{W}:\quad&\bm{\lambda}\eqdef(a_1,a_2,a_3,a_4),\quad
  \text{Re}\,a_i>0\ \ (i=1,\ldots,4),\n
  &\qquad
  \{a_1^*,a_2^*,a_3^*,a_4^*\}=\{a_1,a_2,a_3,a_4\} \quad (\text{as a set}),
  \label{paraW}\\
  \text{AW}:\quad&q^{\bm{\lambda}}\eqdef(a_1,a_2,a_3,a_4),\quad
  |a_i|<1,\ \ (i=1,\ldots,4),\quad 0<q<1,\n
  &\qquad
  \{a_1^*,a_2^*,a_3^*,a_4^*\}=\{a_1,a_2,a_3,a_4\}\ \ (\text{as a set}),
  \label{paraAW}
\end{align}
where
$q^{(\lambda_1,\lambda_2,\ldots)}\eqdef(q^{\lambda_1},q^{\lambda_2},\ldots)$.
The the sinusoidal coordinate $\eta(x)$ is,
\begin{equation}
  \eta(x)\eqdef\left\{
  \begin{array}{lllll}
  x,&x_1=-\infty,& x_2=\infty,&\gamma=1&:\text{cH}\\
  x^2,&x_1=0,&x_2=\infty,&\gamma=1&:\text{W}\\
  \cos x,&x_1=0,& x_2=\pi,&\gamma=\log q&:\text{AW}
  \end{array}\right.\!\!.
  \label{dsinu}
\end{equation}
The potential function $V(x;\bm{\lambda})$ and energy eigenvalue
$\mathcal{E}_n(\bm{\lambda})$ are
\begin{align}
  &V(x;\bm{\lambda})\eqdef\left\{
  \begin{array}{ll}
  (a_1+ix)(a_2+ix)&:\text{cH}\\[2pt]
  \bigl(2ix(2ix+1)\bigr)^{-1}\prod_{j=1}^4(a_j+ix)&:\text{W}\\[2pt]
  \bigl((1-e^{2ix})(1-qe^{2ix})\bigr)^{-1}\prod_{j=1}^4(1-a_je^{ix})
  &:\text{AW}
  \end{array}\right.\!\!,
  \label{Vform}\\
  &\mathcal{E}_n(\bm{\lambda})\eqdef\left\{
  \begin{array}{lll}
  n(n+b_1-1),&b_1\eqdef a_1+a_2+a_1^*+a_2^*&:\text{cH}\\[1pt]
  n(n+b_1-1),&b_1\eqdef a_1+a_2+a_3+a_4&:\text{W}\\[1pt]
  (q^{-n}-1)(1-b_4q^{n-1}),&b_4\eqdef a_1a_2a_3a_4&:\text{AW}
  \end{array}\right.\!\!.
\end{align}
Throughout this paper we consider the potential functions, eigenfunctions,
etc as analytic functions of $x$ in the complex region containing
$x_1<\text{Re}\,x<x_2$.
We use the $*$-operation on an analytic function $*:f\mapsto f^*$
in the following sense.
If $f(x)=\sum\limits_{n}a_nx^n$, $a_n\in\mathbb{C}$,
then $f^*(x)\eqdef\sum\limits_{n}a_n^*x^n$, in which $a_n^*$ is the complex
conjugation of $a_n$. Obviously $f^{**}(x)=f(x)$ and $f(x)^*=f^*(x^*)$.
If a function satisfies $f^*=f$, we call it a `real' function, for it
takes real values on the real line.

The eigenfunctions are chosen real, $\phi_n^*=\phi_n$ and
$\phi_0^*=\phi_0$ and $P_n^*=P_n$. The main part consists of an
orthogonal polynomial $P_n(\eta;\bm{\lambda})$, a polynomial of degree
$n$ in $\eta$:
\begin{equation}
  P_n(\eta;\bm{\lambda})\eqdef\left\{
  \begin{array}{ll}
  p_n(\eta;a_1,a_2,a_1^*,a_2^*)\!\!&:\text{cH}\\[1pt]
  W_n(\eta;a_1,a_2,a_3,a_4)\!\!&:\text{W}\\[1pt]
  p_n(\eta;a_1,a_2,a_3,a_4|q)\!\!&:\text{AW}
  \end{array}\right.\!\!.
\end{equation}
They are expressed in terms of the (basic) hypergeometric functions
\cite{koeswart}:
\begin{align}
  \text{cH:}\quad&p_n(\eta(x);a_1,a_2,a_1^*,a_2^*)\n
  &\quad\eqdef
  i^n\frac{(a_1+a_1^*)_n(a_1+a_2^*)_n}{n!}\,
  {}_3F_2\Bigl(\genfrac{}{}{0pt}{}{-n,\,n+a_1+a_2+a_1^*+a_2^*-1,\,a_1+ix}
  {a_1+a_1^*,\,a_1+a_2^*}\!\Bigm|\!1\Bigr),
 \label{defcH}\\
  \text{W:}\quad&W_n(\eta(x);a_1,a_2,a_3,a_4)\n
  &\quad\eqdef(a_1+a_2)_n(a_1+a_3)_n(a_1+a_4)_n\n
  &\qquad\qquad\times
  {}_4F_3\Bigl(
  \genfrac{}{}{0pt}{}{-n,\,n+\sum_{j=1}^4a_j-1,\,a_1+ix,\,a_1-ix}
  {a_1+a_2,\,a_1+a_3,\,a_1+a_4}\Bigm|1\Bigr),
  \label{defW}\\
  \text{AW:}\quad&p_n(\eta(x);a_1,a_2,a_3,a_4|q)\n
  &\quad\eqdef a_1^{-n}(a_1a_2,a_1a_3,a_1a_4\,;q)_n
  \times
  {}_4\phi_3\Bigl(\genfrac{}{}{0pt}{}{q^{-n},\,a_1a_2a_3a_4q^{n-1},\,
  a_1e^{ix},\,a_1e^{-ix}}{a_1a_2,\,a_1a_3,\,a_1a_4}\!\!\Bigm|\!q\,;q\Bigr),
  \label{defAW}
\end{align}
in which $(a)_n$ and $(a;q)_n$ are the Pochhammer symbol and its $q$-analogue.
They are symmetric in $(a_1,a_2)$ for cH and in $(a_1,a_2,a_3,a_4)$ for
W and AW.

These Hamiltonian systems are exactly solvable in the Schr\"odinger picture.
Shape invariance \cite{genden} is a sufficient condition for the exact
solvability.
Its relations involve one more positive constant $\kappa$:
\begin{align}
  &\mathcal{A}(\bm{\lambda})\mathcal{A}(\bm{\lambda})^{\dagger}
  =\kappa\mathcal{A}(\bm{\lambda+\bm{\delta}})^{\dagger}
  \mathcal{A}(\bm{\lambda}+\bm{\delta})
  +\mathcal{E}_1(\bm{\lambda}),\\
  &\quad\bm{\delta}\eqdef\left\{
  \begin{array}{ll}
  (\frac12,\frac12)&:\text{cH}\\[1pt]
  (\frac12,\frac12,\frac12,\frac12)&:\text{W,\,AW}
  \end{array}\right.\!\!,
  \quad
  \kappa\eqdef\left\{
  \begin{array}{ll}
  1&:\text{cH,\,W}\\
  q^{-1}&:\text{AW}
  \end{array}\right.\!\!,
\end{align}
or equivalently,
\begin{align}
  V(x-i\tfrac{\gamma}{2};\bm{\lambda})
  V^*(x-i\tfrac{\gamma}{2};\bm{\lambda})
  &=\kappa^2\,V(x;\bm{\lambda}+\bm{\delta})
  V^*(x-i\gamma;\bm{\lambda}+\bm{\delta}),\\
  V(x+i\tfrac{\gamma}{2};\bm{\lambda})
  +V^*(x-i\tfrac{\gamma}{2};\bm{\lambda})
  &=\kappa\bigl(V(x;\bm{\lambda}+\bm{\delta})
  +V^*(x;\bm{\lambda}+\bm{\delta})\bigr)
  -\mathcal{E}_1(\bm{\lambda}).
\end{align}
It is straightforward to verify these relations for the above potential
functions for the cH, W and AW cases \eqref{Vform}.
The groundstate wavefunction $\phi_0(x;\bm{\lambda})$ is determined
as a zero mode of the operator $\mathcal{A}(\bm{\lambda})$,
$\mathcal{A}(\bm{\lambda})\phi_0(x;\bm{\lambda})=0$, namely,
\begin{equation}
  \sqrt{V^*(x-i\tfrac{\gamma}{2};\bm{\lambda})}
  \,\phi_0(x-i\tfrac{\gamma}{2};\bm{\lambda})
  =\sqrt{V(x+i\tfrac{\gamma}{2};\bm{\lambda})}
  \,\phi_0(x+i\tfrac{\gamma}{2};\bm{\lambda}),
\end{equation}
and its explicit forms are:
\begin{equation}
  \phi_0(x;\bm{\lambda})\eqdef\left\{
  \begin{array}{ll}
  \sqrt{\Gamma(a_1+ix)\Gamma(a_2+ix)\Gamma(a_1^*-ix)\Gamma(a_2^*-ix)}
  &:\text{cH}\\[2pt]
  \sqrt{(\Gamma(2ix)\Gamma(-2ix))^{-1}\prod_{j=1}^4
  \Gamma(a_j+ix)\Gamma(a_j-ix)}&:\text{W}\\[2pt]
  \sqrt{(e^{2ix}\,;q)_{\infty}(e^{-2ix}\,;q)_{\infty}
  \prod_{j=1}^4(a_je^{ix}\,;q)_{\infty}^{-1}
  (a_je^{-ix}\,;q)_{\infty}^{-1}}
  &:\text{AW}
  \end{array}\right.\!\!.
\end{equation}
We introduce an auxiliary function $\varphi(x)$
\begin{equation}
  \varphi(x)\eqdef\left\{
  \begin{array}{ll}
  1&:\text{cH}\\
  2x&:\text{W}\\
  2\sin x&:\text{AW}
  \end{array}\right.\!\!,
  \label{varphidef}
\end{equation}
which possesses the properties:
\begin{align}
  &\phi_0(x;\bm{\lambda}+\bm{\delta})
  =\varphi(x)\sqrt{V(x+i\tfrac{\gamma}{2};\bm{\lambda})}\,
  \phi_0(x+i\tfrac{\gamma}{2};\bm{\lambda}),\\
  &V(x;\bm{\lambda}+\bm{\delta})
  =\kappa^{-1}\frac{\varphi(x-i\gamma)}{\varphi(x)}
  V(x-i\tfrac{\gamma}{2};\bm{\lambda}).
  \label{varphiprop3}
\end{align}

The action of the operators $\mathcal{A}(\bm{\lambda})$ and
$\mathcal{A}(\bm{\lambda})^{\dagger}$ on the eigenfunctions is
\begin{align}
  &\mathcal{A}(\bm{\lambda})\phi_n(x;\bm{\lambda})
  =f_n(\bm{\lambda})
  \phi_{n-1}\bigl(x;\bm{\lambda}+\bm{\delta}\bigr),
  \label{Aphi=fphi}\\
  &\mathcal{A}(\bm{\lambda})^{\dagger}
  \phi_{n-1}\bigl(x;\bm{\lambda}+\bm{\delta}\bigr)
  =b_{n-1}(\bm{\lambda})\phi_n(x;\bm{\lambda}).
  \label{Adphi=bphi}
\end{align}
The factors of the energy eigenvalue, $f_n(\bm{\lambda})$ and
$b_{n-1}(\bm{\lambda})$,
$\mathcal{E}_n(\bm{\lambda})=f_n(\bm{\lambda})b_{n-1}(\bm{\lambda})$,
are given by
\begin{equation}
  f_n(\bm{\lambda})\eqdef\left\{
  \begin{array}{ll}
  n+b_1-1&:\text{cH}\\
  -n(n+b_1-1)&:\text{W}\\
  q^{\frac{n}{2}}(q^{-n}-1)(1-b_4q^{n-1})&:\text{AW}
  \end{array}\right.\!\!,
  \quad
  b_{n-1}(\bm{\lambda})\eqdef\left\{
  \begin{array}{ll}
  n&:\text{cH}\\
  -1&:\text{W}\\
  q^{-\frac{n}{2}}&:\text{AW}
  \end{array}\right.\!\!.
\end{equation}
The forward and backward shift operators $\mathcal{F}(\bm{\lambda})$ and
$\mathcal{B}(\bm{\lambda})$ are defined by
\begin{align}
  &\mathcal{F}(\bm{\lambda})\eqdef
  \phi_0(x;\bm{\lambda}+\bm{\delta})^{-1}\circ
  \mathcal{A}(\bm{\lambda})\circ\phi_0(x;\bm{\lambda})
  =i\varphi(x)^{-1}(e^{\frac{\gamma}{2}p}-e^{-\frac{\gamma}{2}p}),
  \label{Fdef}\\
  &\mathcal{B}(\bm{\lambda})\eqdef
  \phi_0(x;\bm{\lambda})^{-1}\circ
  \mathcal{A}(\bm{\lambda})^{\dagger}
  \circ\phi_0(x;\bm{\lambda}+\bm{\delta})
  =-i\bigl(V(x;\bm{\lambda})e^{\frac{\gamma}{2}p}
  -V^*(x;\bm{\lambda})e^{-\frac{\gamma}{2}p}\bigr)\varphi(x),
  \label{Bdef}
\end{align}
and their action on the polynomials is
\begin{align}
  &\mathcal{F}(\bm{\lambda})P_n(\eta(x);\bm{\lambda})
  =f_n(\bm{\lambda})P_{n-1}(\eta(x);\bm{\lambda}+\bm{\delta}),
  \label{FP=fP}\\
  &\mathcal{B}(\bm{\lambda})P_{n-1}(\eta(x);\bm{\lambda}+\bm{\delta})
  =b_{n-1}(\bm{\lambda})P_n(\eta(x);\bm{\lambda}).
  \label{BP=bP}
\end{align}
The second order difference operator $\widetilde{\mathcal{H}}(\bm{\lambda})$
acting on the polynomial eigenfunctions is defined by
\begin{align}
  &\widetilde{\mathcal{H}}(\bm{\lambda})\eqdef
  \mathcal{B}(\bm{\lambda})\mathcal{F}(\bm{\lambda})
  =\phi_0(x;\bm{\lambda})^{-1}\circ\mathcal{H}(\bm{\lambda})
  \circ\phi_0(x;\bm{\lambda})\n
  &\phantom{\widetilde{\mathcal{H}}_{\ell}(\bm{\lambda})}
  =V(x;\bm{\lambda})(e^{\gamma p}-1)
  +V^*(x;\bm{\lambda})(e^{-\gamma p}-1),\\
  &\widetilde{\mathcal{H}}(\bm{\lambda})P_n(\eta(x);\bm{\lambda})
  =\mathcal{E}_n(\bm{\lambda})P_n(\eta(x);\bm{\lambda}).
  \label{HtP=EP}
\end{align}
In conventional terms, this is the difference equation determining the
polynomials:
\begin{align}
  &V(x;\bm{\lambda})\bigl(P_n(\eta(x-i\gamma);\bm{\lambda})
  -P_n(\eta(x);\bm{\lambda})\bigr)
  +V^*(x;\bm{\lambda})\bigl(P_n(\eta(x+i\gamma);\bm{\lambda})
  -P_n(\eta(x);\bm{\lambda})\bigr)\n
  &=\mathcal{E}_n(\bm{\lambda})P_n(\eta(x);\bm{\lambda}).
  \label{difeqP}
\end{align}

The orthogonality relation is
\begin{align}
  &\int_{x_1}^{x_2}\!\!\phi_0(x;\bm{\lambda})^2\,
  P_n(\eta(x);\bm{\lambda})P_m(\eta(x);\bm{\lambda})dx
  =h_n(\bm{\lambda})\delta_{nm},
  \label{intPnPm}\\
  &h_n(\bm{\lambda})\eqdef\left\{
  \begin{array}{ll}
  2\pi\prod_{i,j=1}^2\Gamma(n+a_i+a_j^*)\cdot
  \bigl(n!(2n+b_1-1)\Gamma(n+b_1-1)\bigr)^{-1}&:\text{cH}\\[2pt]
  2\pi n!\,(n+b_1-1)_n\prod_{1\leq i<j\leq 4}\Gamma(n+a_i+a_j)\cdot
  \Gamma(2n+b_1)^{-1}&:\text{W}\\[2pt]
  2\pi(b_4q^{n-1};q)_n(b_4q^{2n};q)_{\infty}(q^{n+1};q)_{\infty}^{-1}
  \prod_{1\leq i<j\leq 4}(a_ia_jq^n;q)_{\infty}^{-1}&:\text{AW}
  \end{array}\right.\!\!.
  \label{hn}
\end{align}

As shown in detail in \cite{os7} these Hamiltonian systems are exactly
solvable in the Heisenberg picture, too.
The positive/negative frequency parts of the exact Heisenberg operator
solution of the sinusoidal coordinate $\eta(x)$ provide the
{\em annihilation/creation\/} operators $a^{(\pm)}(\bm{\lambda})$ which
map the eigenfunctions to the neighbouring levels
\begin{equation}
  a^{(\pm)}(\bm{\lambda})\phi_n(x;\bm{\lambda})\propto
  \phi_{n\pm1}(x;\bm{\lambda}).
  \label{anncreori}
\end{equation}
This is a disguise of the three term recurrence relations of these
orthogonal polynomials.
The above relations are to be contrasted with the actions of the
operators $A(\bm{\lambda})$ and $A(\bm{\lambda}-\bm{\delta})^\dagger$
\eqref{Aphi=fphi}--\eqref{Adphi=bphi}, which maps $\phi_n(x;\bm{\lambda})$
to the neighbouring levels of {\em shifted parameters\/}
$\bm{\lambda}\pm\bm{\delta}$.

\subsection{The deformed systems}
\label{sec:deformed.sys}

Here we recapitulate the Hamiltonian systems of the exceptional Wilson and
Askey-Wilson polynomials, which were derived by the present authors
\cite{os17} in 2009.
The exceptional continuous Hahn polynomials are new.
For each $\ell=1,2,\ldots$, we can construct a shape invariant system
by deforming the original system ($\ell=0$) in terms of a degree $\ell$
eigenpolynomial $\xi_{\ell}(\eta)$ of twisted parameters.
We restrict the original parameter ranges \eqref{paracH}--\eqref{paraAW}
as follows:
\begin{align}
  \text{cH}:\quad&a_1>0,\quad\text{$\ell$ : even},
  \label{paracHrest}\\
  \text{W}:\quad&a_1,a_2\in\mathbb{R},\quad
  \{a_3^*,a_4^*\}=\{a_3,a_4\} \quad (\text{as a set}),\n
  &0<a_j<\text{Re}\,a_k\ \ (j=1,2;k=3,4),
  \label{paraWrest}\\
  \text{AW}:\quad&a_1,a_2\in\mathbb{R},\quad
  \{a_3^*,a_4^*\}=\{a_3,a_4\} \quad (\text{as a set}),\n
  &1>a_j>|a_k|\ \ (j=1,2;k=3,4).
  \label{paraAWrest}
\end{align}
In terms of the twist operation $\mathfrak{t}$ acting on the set of
parameters,
\begin{equation}
  \mathfrak{t}(\bm{\lambda})\eqdef\left\{
  \begin{array}{ll}
  (-\lambda_1,\lambda_2)&:\text{cH}\\
  (-\lambda_1,-\lambda_2,\lambda_3,\lambda_4)&:\text{W,\,AW}
  \end{array}\right.\!\!,
\end{equation}
the deforming polynomial $\xi_{\ell}(\eta)$ is defined from the
eigenpolynomial $P_{\ell}(\eta)$:
\begin{equation}
  \xi_{\ell}(\eta;\bm{\lambda})\eqdef
  P_{\ell}\bigl(\eta;\mathfrak{t}
  \bigl(\bm{\lambda}+(\ell-1)\bm{\delta}\bigr)\bigr).
\end{equation}
We may need to restrict parameters further in order that
$\xi_{\ell}(\eta(x);\bm{\lambda})$ has no zero in the rectangular domain
$x_1\leq\text{Re}\,x\leq x_2$, $|\text{Im}\,x|\leq\frac12|\gamma|$,
which is necessary for the hermiticity of the Hamiltonian.
The potential function, the Hamiltonian and the Schr\"{o}dinger equation
of the deformed system are:
\begin{align}
  &V_{\ell}(x;\bm{\lambda})\eqdef
  V(x;\bm{\lambda}+\ell\bm{\delta})\,
  \frac{\xi_{\ell}(\eta(x+i\frac{\gamma}{2});\bm{\lambda})}
  {\xi_{\ell}(\eta(x-i\frac{\gamma}{2});\bm{\lambda})}
  \frac{\xi_{\ell}(\eta(x-i\gamma);\bm{\lambda}+\bm{\delta})}
  {\xi_{\ell}(\eta(x);\bm{\lambda}+\bm{\delta})},
  \label{Vl}\\
  &V_{\ell}^*(x;\bm{\lambda})=
  V^*(x;\bm{\lambda}+\ell\bm{\delta})\,
  \frac{\xi_{\ell}(\eta(x-i\frac{\gamma}{2});\bm{\lambda})}
  {\xi_{\ell}(\eta(x+i\frac{\gamma}{2});\bm{\lambda})}
  \frac{\xi_{\ell}(\eta(x+i\gamma);\bm{\lambda}+\bm{\delta})}
  {\xi_{\ell}(\eta(x);\bm{\lambda}+\bm{\delta})},
  \label{Vl*}\\
  &\mathcal{A}_{\ell}(\bm{\lambda})\eqdef
   i\bigl(e^{\frac{\gamma}{2}p}\sqrt{V_{\ell}^*(x;\bm{\lambda})}
  -e^{-\frac{\gamma}{2}p}\sqrt{V_{\ell}(x;\bm{\lambda})}\,\bigr),\n
  &\mathcal{A}_{\ell}(\bm{\lambda})^{\dagger}\eqdef
   -i\bigl(\sqrt{V_{\ell}(x;\bm{\lambda})}\,e^{\frac{\gamma}{2}p}
  -\sqrt{V_{\ell}^*(x;\bm{\lambda})}\,e^{-\frac{\gamma}{2}p}\bigr),\\
  &\mathcal{H}_{\ell}(\bm{\lambda})\eqdef
  \mathcal{A}_{\ell}(\bm{\lambda})^{\dagger}
  \mathcal{A}_{\ell}(\bm{\lambda}),
  \label{deformham}\\
  &\mathcal{H}_{\ell}(\bm{\lambda})\phi_{\ell,n}(x;\bm{\lambda})
  =\mathcal{E}_{\ell,n}(\bm{\lambda})\phi_{\ell,n}(x;\bm{\lambda})\quad
  (n=0,1,2,\ldots),\quad
  \mathcal{E}_{\ell,n}(\bm{\lambda})
  =\mathcal{E}_n(\bm{\lambda}+\ell\bm{\delta}).
\end{align}
The continuous Hahn polynomials are real polynomials defined on the
entire real line.
Therefore the odd degree members have at least one real zero, whichever
coefficients we may choose. This is the reason why $\ell$ is restricted
to even integers in the cH case \eqref{paracHrest}. 

This system is shape invariant:
\begin{equation}
  \mathcal{A}_{\ell}(\bm{\lambda})\mathcal{A}_{\ell}(\bm{\lambda})^{\dagger}
  =\kappa\mathcal{A}_{\ell}(\bm{\lambda+\bm{\delta}})^{\dagger}
  \mathcal{A}_{\ell}(\bm{\lambda}+\bm{\delta})
  +\mathcal{E}_{\ell,1}(\bm{\lambda}),
\end{equation}
or equivalently,
\begin{align}
  V_{\ell}(x-i\tfrac{\gamma}{2};\bm{\lambda})
  V_{\ell}^*(x-i\tfrac{\gamma}{2};\bm{\lambda})
  &=\kappa^2\,V_{\ell}(x;\bm{\lambda}+\bm{\delta})
  V_{\ell}^*(x-i\gamma;\bm{\lambda}+\bm{\delta}),
  \label{shapeinvVV}\\
  V_{\ell}(x+i\tfrac{\gamma}{2};\bm{\lambda})
  +V_{\ell}^*(x-i\tfrac{\gamma}{2};\bm{\lambda})
  &=\kappa\bigl(V_{\ell}(x;\bm{\lambda}+\bm{\delta})
  +V_{\ell}^*(x;\bm{\lambda}+\bm{\delta})\bigr)
  -\mathcal{E}_{\ell,1}(\bm{\lambda}).
  \label{shapeinvV}
\end{align}
Proof is straightforward by direct calculation. In order to derive
eq.\,\eqref{shapeinvV}, use is made of the two properties of the
deforming polynomial $\xi_{\ell}$ \eqref{xil(l+d)}--\eqref{xil(l)}
and the factorisation of the potential \eqref{factorV}.
For another simple proof of shape invariance, see the discussion
in the final section.
The eigenfunctions are
\begin{align}
  &\psi_{\ell}(x;\bm{\lambda})
  \eqdef\frac{\phi_0(x;\bm{\lambda}+\ell\bm{\delta})}
  {\sqrt{\xi_{\ell}(\eta(x+i\frac{\gamma}{2});\bm{\lambda})
  \xi_{\ell}(\eta(x-i\frac{\gamma}{2});\bm{\lambda})}}\,,
  \\[2pt]
  &\phi_{\ell,n}(x;\bm{\lambda})=\psi_{\ell}(x;\bm{\lambda})
  P_{\ell,n}\bigl(\eta(x);\bm{\lambda}\bigr)\quad
  (n=0,1,2,\ldots).
\end{align}
Here $P_{\ell,n}(\eta;\bm{\lambda})$ is a degree $\ell+n$ polynomial
in $\eta$ but $P_{\ell,n}(\eta(x);\bm{\lambda})$ has only $n$ zeros
in the domain $x_1<x<x_2$.
The explicit forms of $P_{\ell,n}(\eta)$ were given by eqs.\,(42)--(44) in
\cite{os17}, with eqs.\,(66)-(68) for W and eqs.\,(80)--(82) for AW.
Here we present much simpler looking forms of them, which encompasses
the new cH case, too:
\begin{align}
  P_{\ell,n}(\eta(x);\bm{\lambda})
  &=\frac{-i}{\hat{f}_{\ell,n}(\bm{\lambda})\varphi(x)}\Bigl(
  v_1(x;\bm{\lambda}+\ell\bm{\delta})
  \xi_{\ell}(\eta(x+i\tfrac{\gamma}{2});\bm{\lambda})
  P_n(\eta(x-i\tfrac{\gamma}{2});\bm{\lambda}+\ell\bm{\delta}
  +\bm{\tilde{\delta}})\n
  &\qquad\qquad
  -v_1^*(x;\bm{\lambda}+\ell\bm{\delta})
  \xi_{\ell}(\eta(x-i\tfrac{\gamma}{2});\bm{\lambda})
  P_n(\eta(x+i\tfrac{\gamma}{2});\bm{\lambda}+\ell\bm{\delta}
  +\bm{\tilde{\delta}})\Bigr),
  \label{mainres}
\end{align}
where $v_1(x;\bm{\lambda})$, $\hat{f}_{\ell,n}(\bm{\lambda})$ and
$\bm{\tilde{\delta}}$ will be defined in \eqref{v1v2}, \eqref{hatfhatb}
and \eqref{hatkappa}.
This is one of the main results of the present paper which  is derived
in \S\,\ref{sec:intertwin}.
Its lowest degree member is the degree $\ell$ deforming polynomial
itself of the shifted parameters
\begin{equation}
  P_{\ell,0}(\eta(x);\bm{\lambda})
  =\xi_{\ell}(\eta(x);\bm{\lambda}+\bm{\delta}),
\end{equation}
which is obtained by \eqref{xil(l+d)}. It is straightforward to verify
that the groundstate eigenfunction
\begin{equation}
  \phi_{\ell,0}(x;\bm{\lambda})
  =\frac{\phi_0(x;\bm{\lambda}+\ell\bm{\delta})\,
  \xi_{\ell}(\eta(x);\bm{\lambda}+\bm{\delta})}
  {\sqrt{\xi_{\ell}(\eta(x+i\frac{\gamma}{2});\bm{\lambda})
  \xi_{\ell}(\eta(x-i\frac{\gamma}{2});\bm{\lambda})}}
\end{equation}
is the zero mode of the operator $\mathcal{A}_{\ell}(\bm{\lambda})$,
$\mathcal{A}_{\ell}(\bm{\lambda})\phi_{\ell,0}(x;\bm{\lambda})=0$.

The action of $\mathcal{A}_{\ell}(\bm{\lambda})$ and
$\mathcal{A}_{\ell}(\bm{\lambda})^{\dagger}$ on the eigenfunctions is
\begin{align}
  &\mathcal{A}_{\ell}(\bm{\lambda})\phi_{\ell,n}(x;\bm{\lambda})
  =f_{\ell,n}(\bm{\lambda})
  \phi_{\ell,n-1}\bigl(x;\bm{\lambda}+\bm{\delta}\bigr),
  \label{Alphiln=flnphiln}\\
  &\mathcal{A}_{\ell}(\bm{\lambda})^{\dagger}
  \phi_{\ell,n-1}\bigl(x;\bm{\lambda}+\bm{\delta}\bigr)
  =b_{\ell,n-1}(\bm{\lambda})\phi_{\ell,n}(x;\bm{\lambda}),
  \label{Aldphiln=blnphiln}\\
  &f_{\ell,n}(\bm{\lambda})=f_n(\bm{\lambda}+\ell\bm{\delta}),\quad
  b_{\ell,n-1}(\bm{\lambda})=b_{n-1}(\bm{\lambda}+\ell\bm{\delta}).
  \label{fln,bln}
\end{align}
Like the corresponding formulas of the original systems
\eqref{Aphi=fphi}--\eqref{Adphi=bphi}, these are simple consequences of
the shape invariance and the normalisation of the eigenfunctions.
In the next section, we will derive these formulas through the
intertwining relations and without assuming shape invariance.

The forward shift operator $\mathcal{F}_{\ell}(\bm{\lambda})$ and
the backward shift operator $\mathcal{B}_{\ell}(\bm{\lambda})$ are
defined in a similar way as before
\begin{align}
  \mathcal{F}_{\ell}(\bm{\lambda})&\eqdef
  \psi_{\ell}(x;\bm{\lambda}+\bm{\delta})^{-1}\circ
  \mathcal{A}_{\ell}(\bm{\lambda})\circ
  \psi_{\ell}(x;\bm{\lambda})\n
  &=\frac{i}{\varphi(x)\xi_{\ell}(\eta(x);\bm{\lambda})}
  \Bigl(\xi_{\ell}\bigl(\eta(x+i\tfrac{\gamma}{2});\bm{\lambda}
  +\bm{\delta}\bigr)e^{\frac{\gamma}{2}p}
  -\xi_{\ell}\bigl(\eta(x-i\tfrac{\gamma}{2});\bm{\lambda}
  +\bm{\delta}\bigr)e^{-\frac{\gamma}{2}p}\Bigr),
  \label{Fldef}\\
  \mathcal{B}_{\ell}(\bm{\lambda})&\eqdef
  \psi_{\ell}(x;\bm{\lambda})^{-1}\circ
  \mathcal{A}_{\ell}(\bm{\lambda})^{\dagger}\circ
  \psi_{\ell}(x;\bm{\lambda}+\bm{\delta})\n
  &=\frac{-i}{\xi_{\ell}(\eta(x);\bm{\lambda}+\bm{\delta})}
  \Bigl(V(x;\bm{\lambda}+\ell\bm{\delta})
  \xi_{\ell}\bigl(\eta(x+i\tfrac{\gamma}{2});\bm{\lambda}\bigr)
  e^{\frac{\gamma}{2}p}\n
  &\qquad\qquad\qquad\qquad
  -V^*(x;\bm{\lambda}+\ell\bm{\delta})
  \xi_{\ell}\bigl(\eta(x-i\tfrac{\gamma}{2});\bm{\lambda}\bigr)
  e^{-\frac{\gamma}{2}p}\Bigr)\varphi(x),
  \label{Bldef}
\end{align}
and their action on the polynomial $P_{\ell,n}(\eta;\bm{\lambda})$ is
\begin{align}
  &\mathcal{F}_{\ell}(\bm{\lambda})
  P_{\ell,n}\bigl(\eta(x);\bm{\lambda}\bigr)
  =f_{\ell,n}(\bm{\lambda})
  P_{\ell,n-1}\bigl(\eta(x);\bm{\lambda}+\bm{\delta}\bigr),
  \label{FlPln=flnPln}\\
  &\mathcal{B}_{\ell}(\bm{\lambda})
  P_{\ell,n-1}\bigl(\eta(x);\bm{\lambda}+\bm{\delta}\bigr)
  =b_{\ell,n-1}(\bm{\lambda})P_{\ell,n}\bigl(\eta(x);\bm{\lambda}\bigr).
  \label{BlPln=blnPln}
\end{align}
The second order difference operator
$\widetilde{\mathcal{H}}_{\ell}(\bm{\lambda})$ acting on the polynomial
eigenfunctions is defined by
\begin{align}
  &\widetilde{\mathcal{H}}_{\ell}(\bm{\lambda})\eqdef
  \mathcal{B}_{\ell}(\bm{\lambda})\mathcal{F}_{\ell}(\bm{\lambda})
  =\psi_{\ell}(x;\bm{\lambda})^{-1}\circ\mathcal{H}_{\ell}(\bm{\lambda})
  \circ\psi_{\ell}(x;\bm{\lambda})\n
  &\phantom{\widetilde{\mathcal{H}}_{\ell}(\bm{\lambda})}
  =V(x;\bm{\lambda}+\ell\bm{\delta})
  \frac{ \xi_{\ell}\bigl(\eta(x+i\tfrac{\gamma}{2});\bm{\lambda}\bigr)}
  {\xi_{\ell}\bigl(\eta(x-i\tfrac{\gamma}{2});\bm{\lambda}\bigr)}
  \biggl(e^{\gamma p}
  -\frac{\xi_{\ell}\bigl(\eta(x-i\gamma);\bm{\lambda}+\bm{\delta}\bigr)}
  {\xi_{\ell}\bigl(\eta(x);\bm{\lambda}+\bm{\delta}\bigr)}\biggr)\n
  &\phantom{\widetilde{\mathcal{H}}_{\ell}(\bm{\lambda})}
  \phantom{=}+V^*(x;\bm{\lambda}+\ell\bm{\delta})
  \frac{ \xi_{\ell}\bigl(\eta(x-i\tfrac{\gamma}{2});\bm{\lambda}\bigr)}
  {\xi_{\ell}\bigl(\eta(x+i\tfrac{\gamma}{2});\bm{\lambda}\bigr)}
  \biggl(e^{-\gamma p}
  -\frac{\xi_{\ell}\bigl(\eta(x+i\gamma);\bm{\lambda}+\bm{\delta}\bigr)}
  {\xi_{\ell}\bigl(\eta(x);\bm{\lambda}+\bm{\delta}\bigr)}\biggr),\\
  &\widetilde{\mathcal{H}}_{\ell}(\bm{\lambda})
  P_{\ell,n}(\eta(x);\bm{\lambda})
  =\mathcal{E}_{\ell,n}(\bm{\lambda})
  P_{\ell,n}(\eta(x);\bm{\lambda}).
\end{align}
Again it is trivial to verify that the lowest degree polynomial
$P_{\ell,0}(\eta(x);\bm{\lambda})
=\xi_{\ell}\bigl(\eta(x);\bm{\lambda}+\bm{\delta}\bigr)$ is the zero
mode of $\widetilde{\mathcal{H}}_{\ell}(\bm{\lambda})$:
\begin{equation}
  \widetilde{\mathcal{H}}_{\ell}(\bm{\lambda})
  \xi_{\ell}\bigl(\eta(x);\bm{\lambda}+\bm{\delta}\bigr)=0.
\end{equation}

The orthogonality relation is
\begin{align}
  &\int_{x_1}^{x_2}\!\!\psi_{\ell}(x;\bm{\lambda})^2\,
  P_{\ell,n}(\eta(x);\bm{\lambda})P_{\ell,m}(\eta(x);\bm{\lambda})dx
  =h_{\ell,n}(\bm{\lambda})\delta_{nm},\\
  &h_{\ell,n}(\bm{\lambda})\eqdef
  h_n(\bm{\lambda}+\ell\bm{\delta})\times\left\{
  \begin{array}{ll}
  {\displaystyle
  \frac{(2a_1+n+\ell)(a_2+a_2^*+n+2\ell-1)}{(2a_1+n)(a_2+a_2^*+n+\ell-1)}}
  &:\text{cH}\\[10pt]
  {\displaystyle
  \frac{(a_1+a_2+n+\ell)(a_3+a_4+n+2\ell-1)}{(a_1+a_2+n)(a_3+a_4+n+\ell-1)}}
  &:\text{W}\\[10pt]
  {\displaystyle
  q^{-\ell}\frac{(1-a_1a_2q^{n+\ell})(1-a_3a_4q^{n+2\ell-1})}
  {(1-a_1a_2q^n)(1-a_3a_4q^{n+\ell-1})}}
  &:\text{AW}
  \end{array}\right.\!\!.
  \label{hln}
\end{align}

\subsection{Properties of the deforming polynomial $\xi_{\ell}$}
\label{sec:xil_prop}

Here we present three formulas of the deforming polynomial
$\xi_{\ell}(\eta;\bm{\lambda})$ \eqref{xildiffeq}--\eqref{xil(l)},
which will play important roles in the derivation of various results
in section three, in particular, the fundamental results of this paper
\eqref{Hl+=H} and \eqref{Hl-=Hl}:
\begin{align}
  &\Bigl(V\bigl(x;\mathfrak{t}(\bm{\lambda}+(\ell-1)\bm{\delta})\bigr)
  (e^{\gamma p}-1)
  +V^*\bigl(x;\mathfrak{t}(\bm{\lambda}+(\ell-1)\bm{\delta})\bigr)
  (e^{-\gamma p}-1)\Bigr)\xi_{\ell}(\eta(x);\bm{\lambda})\n
  &\qquad\quad
  =\mathcal{E}_{\ell}(\mathfrak{t}(\bm{\lambda}))
  \xi_{\ell}(\eta(x);\bm{\lambda}),
  \label{xildiffeq}\\
  &\tfrac{i}{\varphi(x)}\bigl(
  v_1^*(x;\bm{\lambda}+\ell\bm{\delta})e^{\frac{\gamma}{2}p}
  -v_1(x;\bm{\lambda}+\ell\bm{\delta})e^{-\frac{\gamma}{2}p}\bigr)
  \xi_{\ell}(\eta(x);\bm{\lambda})
  =\hat{f}_{\ell,0}(\bm{\lambda})\xi_{\ell}(\eta(x);\bm{\lambda}+\bm{\delta}),
  \label{xil(l+d)}\\
  &\tfrac{-i}{\varphi(x)}\bigl(
  v_2(x;\bm{\lambda}+(\ell-1)\bm{\delta})e^{\frac{\gamma}{2}p}
  -v_2^*(x;\bm{\lambda}+(\ell-1)\bm{\delta})e^{-\frac{\gamma}{2}p}\bigr)
  \xi_{\ell}(\eta(x);\bm{\lambda}+\bm{\delta})\n
  &\qquad\quad
  =\hat{b}_{\ell,0}(\bm{\lambda})\xi_{\ell}(\eta(x);\bm{\lambda}),
  \label{xil(l)}
\end{align}
where $v_1(x;\bm{\lambda})$, $v_2(x;\bm{\lambda})$ are the factors of
the potential function $V(x;\bm{\lambda})$:
\begin{align}
  &\hspace{40mm}V(x;\bm{\lambda})=-\sqrt{\kappa}\,
  \frac{v_1(x;\bm{\lambda})v_2(x;\bm{\lambda})}
  {\varphi(x)\varphi(x-i\frac{\gamma}{2})},
  \label{factorV}\\
  &\!\!\!v_1(x;\bm{\lambda})\eqdef\left\{
  \begin{array}{ll}
  \!\!i(a_1+ix)&\!\!:\text{cH}\\
  \!\!\prod_{j=1}^2(a_j+ix)&\!\!:\text{W}\\
  \!\!e^{-ix}\prod_{j=1}^2(1-a_je^{ix})&\!\!:\text{AW}
  \end{array}\right.\!\!\!,
  \ \ v_2(x;\bm{\lambda})\eqdef\left\{
  \begin{array}{ll}
  \!\!i(a_2+ix)&\!\!:\text{cH}\\
  \!\!\prod_{j=3}^4(a_j+ix)&\!\!:\text{W}\\
  \!\!e^{-ix}\prod_{j=3}^4(1-a_je^{ix})&\!\!:\text{AW}
  \end{array}\right.\!\!\!.\!\!\!
  \label{v1v2}
\end{align}
The constants $\hat{f}_{\ell,n}(\bm{\lambda})$ and
$\hat{b}_{\ell,n}(\bm{\lambda})$ are given by
\begin{equation}
  \hat{f}_{\ell,n}(\bm{\lambda})\eqdef\left\{
  \begin{array}{ll}
  \!\!2a_1+n&\!\!\!:\text{cH}\\
  \!\!a_1+a_2+n&\!\!\!:\text{W}\\
  \!\!-q^{-\frac{n-\ell}{2}}(1-a_1a_2q^n)&\!\!\!:\text{AW}
  \end{array}\right.\!\!\!,
  \ \ \hat{b}_{\ell,n}(\bm{\lambda})\eqdef\left\{
  \begin{array}{ll}
  \!\!a_2+a_2^*+n+2\ell-1&\!\!\!:\text{cH}\\
  \!\!a_3+a_4+n+2\ell-1&\!\!\!:\text{W}\\
  \!\!-q^{-\frac{n+\ell}{2}}(1-a_3a_4q^{n+2\ell-1})&\!\!\!:\text{AW}
  \end{array}\right.\!\!\!.\!\!\!
  \label{hatfhatb}
\end{equation}
The first equation \eqref{xildiffeq} is the difference equation for
the deforming polynomial, which corresponds to \eqref{difeqP}.
The eqs.\,\eqref{xil(l+d)}--\eqref{xil(l)} are identities relating
$\xi_{\ell}(\eta;\bm{\lambda})$ and
$\xi_{\ell}(\eta;\bm{\lambda}+\bm{\delta})$.
We remark that these two equations \eqref{xil(l+d)}--\eqref{xil(l)}
imply \eqref{xildiffeq}.
In similar problems in ordinary quantum mechanics, the exceptional
Laguerre and Jacobi polynomials, analogous identities play important
roles in proving shape invariance and other relations \cite{os18,hos,stz}.
As shown in \eqref{defcH}--\eqref{defAW}, the continuous Hahn, Wilson
and Askey-Wilson polynomials are expressed in terms of the (basic)
hypergeometric functions ${}_3F_2$, ${}_4F_3$ and ${}_4\phi_3$,
respectively \cite{koeswart}.
The identities\,\eqref{xil(l+d)}--\eqref{xil(l)} are reduced to the
following identities satisfied by the (basic) hypergeometric functions:
($\alpha$, $\alpha'$, $\alpha_1,\ldots,\alpha_4$ : generic parameters)
\begin{align}
  &\quad
  (\alpha_1+ix)\ {}_3F_2\Bigl(\genfrac{}{}{0pt}{}
  {-\alpha,\alpha',\alpha_1+1+ix}
  {\alpha_3,\alpha_1+\alpha_2+1}\Bigm|1\Bigr)
  +(\alpha_2-ix)\ {}_3F_2\Bigl(\genfrac{}{}{0pt}{}
  {-\alpha,\alpha',\alpha_1+ix}
  {\alpha_3,\alpha_1+\alpha_2+1}\Bigm|1\Bigr)\n
  &=(\alpha_1+\alpha_2)\ {}_3F_2\Bigl(\genfrac{}{}{0pt}{}
  {-\alpha,\alpha',\alpha_1+ix}
  {\alpha_3,\alpha_1+\alpha_2}\Bigm|1\Bigr),
  \label{3F2propB}\\
  &\quad\frac{-i}{2x}\Bigl(
  (\alpha_1+ix)(\alpha_2+ix)\ {}_4F_3\Bigl(\genfrac{}{}{0pt}{}
  {-\alpha,\alpha',\alpha_1+1+ix,\alpha_1-ix}
  {\alpha_1+\alpha_2+1,\alpha_1+\alpha_3,\alpha_1+\alpha_4}\Bigm|1\Bigr)\n
  &\qquad
  -(\alpha_1-ix)(\alpha_2-ix)\ {}_4F_3\Bigl(\genfrac{}{}{0pt}{}
  {-\alpha,\alpha',\alpha_1+ix,\alpha_1+1-ix}
  {\alpha_1+\alpha_2+1,\alpha_1+\alpha_3,\alpha_1+\alpha_4}\Bigm|1\Bigr)
  \Bigr)\n
  &=(\alpha_1+\alpha_2)\ {}_4F_3\Bigl(\genfrac{}{}{0pt}{}
  {-\alpha,\alpha',\alpha_1+ix,\alpha_1-ix}
  {\alpha_1+\alpha_2,\alpha_1+\alpha_3,\alpha_1+\alpha_4}\Bigm|1\Bigr),
  \label{4F3propB}\\
  &\quad\frac{-i}{2\sin x}\Bigl(
  e^{-ix}(1-\alpha_1e^{ix})(1-\alpha_2e^{ix})
  \ {}_4\phi_3\Bigl(\genfrac{}{}{0pt}{}
  {\alpha^{-1},\alpha',\alpha_1qe^{ix},\alpha_1e^{-ix}}
  {\alpha_1\alpha_2q,\alpha_1\alpha_3,\alpha_1\alpha_4}\Bigm|q;q\Bigr)\n
  &\qquad\quad
  -e^{ix}(1-\alpha_1e^{-ix})(1-\alpha_2e^{-ix})
  \ {}_4\phi_3\Bigl(\genfrac{}{}{0pt}{}
  {\alpha^{-1},\alpha',\alpha_1e^{ix},\alpha_1qe^{-ix}}
  {\alpha_1\alpha_2q,\alpha_1\alpha_3,\alpha_1\alpha_4}\Bigm|q;q\Bigr)
  \Bigr)\n
  &=-(1-\alpha_1\alpha_2)\ {}_4\phi_3\Bigl(\genfrac{}{}{0pt}{}
  {\alpha^{-1},\alpha',\alpha_1e^{ix},\alpha_1e^{-ix}}
  {\alpha_1\alpha_2,\alpha_1\alpha_3,\alpha_1\alpha_4}\Bigm|q;q\Bigr).
  \label{4phi3propB}
\end{align}
These identities can be easily verified by the series definition of the
(basic) hypergeometric functions.

\section{Intertwining relations}
\setcounter{equation}{0}
\label{sec:intertwin}

Here we demonstrate that the Hamiltonian system of the original polynomials 
reviewed in \S\,\ref{sec:org.sys} and the deformation summarised in
\S\,\ref{sec:deformed.sys} are intertwined by a discrete version of the
Darboux-Crum transformation.
This provides simple expressions of the eigenfunctions of the deformed
systems \eqref{mainres} in terms of those of the original system,
which is exactly solvable.
It also delivers a simple proof of the shape invariance of the deformed
system.

\subsection{General setting}
\label{sec:general}

For well-defined operators $\hat{\mathcal{A}}_{\ell}(\bm{\lambda})$ and
$\hat{\mathcal{A}}_{\ell}(\bm{\lambda})^{\dagger}$, let us define
a pair of Hamiltonians $\hat{\mathcal{H}}_{\ell}^{(\pm)}(\bm{\lambda})$
\begin{equation}
  \hat{\mathcal{H}}_{\ell}^{(+)}(\bm{\lambda})\eqdef
  \hat{\mathcal{A}}_{\ell}(\bm{\lambda})^{\dagger}
  \hat{\mathcal{A}}_{\ell}(\bm{\lambda}),\quad
  \hat{\mathcal{H}}_{\ell}^{(-)}(\bm{\lambda})\eqdef
  \hat{\mathcal{A}}_{\ell}(\bm{\lambda})
  \hat{\mathcal{A}}_{\ell}(\bm{\lambda})^{\dagger},
  \label{H+-}
\end{equation}
and consider their the Schr\"{o}dinger equations, that is, the
eigenvalue problems: 
\begin{equation}
  \hat{\mathcal{H}}_{\ell}^{(\pm)}(\bm{\lambda})
  \hat{\phi}_{\ell,n}^{(\pm)}(x;\bm{\lambda})
  =\hat{\mathcal{E}}_{\ell,n}^{(\pm)}(\bm{\lambda})
  \hat{\phi}_{\ell,n}^{(\pm)}(x;\bm{\lambda})\quad
  (n=0,1,2,\ldots).
  \label{H+-Scheq}
\end{equation}
By definition, all the eigenfunctions must be square integrable.
Obviously the pair of Hamiltonians are intertwined:
\begin{align}
  &\hat{\mathcal{H}}_{\ell}^{(+)}(\bm{\lambda})
  \hat{\mathcal{A}}_{\ell}(\bm{\lambda})^{\dagger}
  =\hat{\mathcal{A}}_{\ell}(\bm{\lambda})^{\dagger}
  \hat{\mathcal{A}}_{\ell}(\bm{\lambda})
  \hat{\mathcal{A}}_{\ell}(\bm{\lambda})^{\dagger}
  =\hat{\mathcal{A}}_{\ell}(\bm{\lambda})^{\dagger}
  \hat{\mathcal{H}}_{\ell}^{(-)}(\bm{\lambda}),\\
  &\hat{\mathcal{A}}_{\ell}(\bm{\lambda})
  \hat{\mathcal{H}}_{\ell}^{(+)}(\bm{\lambda})
  =\hat{\mathcal{A}}_{\ell}(\bm{\lambda})
  \hat{\mathcal{A}}_{\ell}(\bm{\lambda})^{\dagger}
  \hat{\mathcal{A}}_{\ell}(\bm{\lambda})
  =\hat{\mathcal{H}}_{\ell}^{(-)}(\bm{\lambda})
  \hat{\mathcal{A}}_{\ell}(\bm{\lambda}).
\end{align}
If $\hat{\mathcal{A}}_{\ell}(\bm{\lambda})
\hat{\phi}_{\ell,n}^{(+)}(x;\bm{\lambda})\neq 0$ and
$\hat{\mathcal{A}}_{\ell}(\bm{\lambda})^{\dagger}
\hat{\phi}_{\ell,n}^{(-)}(x;\bm{\lambda})\neq 0$, then the two systems are
exactly iso-spectral and there is one-to-one correspondence between the
eigenfunctions:
\begin{align}
  &\hat{\mathcal{E}}_{\ell,n}^{(+)}(\bm{\lambda})
  =\hat{\mathcal{E}}_{\ell,n}^{(-)}(\bm{\lambda}),\\
  &\hat{\phi}_{\ell,n}^{(-)}(x;\bm{\lambda})\propto
  \hat{\mathcal{A}}_{\ell}(\bm{\lambda})
  \hat{\phi}_{\ell,n}^{(+)}(x;\bm{\lambda}),\quad
  \hat{\phi}_{\ell,n}^{(+)}(x;\bm{\lambda})\propto
  \hat{\mathcal{A}}_{\ell}(\bm{\lambda})^{\dagger}
  \hat{\phi}_{\ell,n}^{(-)}(x;\bm{\lambda}).
\end{align}
It should be stressed in the ordinary setting of Crum's theorem, the zero
mode of $ \hat{\mathcal{A}}_{\ell}(\bm{\lambda})$ is the groundstate of
$\hat{\mathcal{H}}_{\ell}^{(+)}(\bm{\lambda})$.
In that case, $\hat{\mathcal{H}}_{\ell}^{(+)}(\bm{\lambda})$ and
$\hat{\mathcal{H}}_{\ell}^{(-)}(\bm{\lambda})$ are iso-spectral except
for the groundstate (essentially iso-spectral) of
$\hat{\mathcal{H}}_{\ell}^{(+)}(\bm{\lambda})$, since it is annihilated by
$\hat{\mathcal{A}}_{\ell}(\bm{\lambda})$ \cite{junkroy}.

In the following we will present the explicit forms of the operators
$\hat{\mathcal{A}}_{\ell}(\bm{\lambda})$ and
$\hat{\mathcal{A}}_{\ell}(\bm{\lambda})^{\dagger}$,
which intertwine the original systems in \S\,\ref{sec:org.sys} and
the deformed systems in \S\,\ref{sec:deformed.sys}.

\subsection{Intertwining the original and the deformed systems of
the polynomials}
\label{sec:intrel}

The potential function $\hat{V}_{\ell}$ is the original potential function
$V$ with {\em twisted parameters\/} and multiplicatively deformed by the
deforming polynomial $\xi_{\ell}$:
\begin{align}
  \hat{V}_{\ell}(x;\bm{\lambda})&\eqdef
  V\bigl(x;\mathfrak{t}(\bm{\lambda}+(\ell-1)\bm{\delta})\bigr)
  \frac{\xi_{\ell}(\eta(x-i\gamma);\bm{\lambda})}
  {\xi_{\ell}(\eta(x);\bm{\lambda})},
  \label{intV}\\
  \hat{V}_{\ell}^*(x;\bm{\lambda})&=
  V^*\bigl(x;\mathfrak{t}(\bm{\lambda}+(\ell-1)\bm{\delta})\bigr)
  \frac{\xi_{\ell}(\eta(x+i\gamma);\bm{\lambda})}
  {\xi_{\ell}(\eta(x);\bm{\lambda})},
  \label{intVs}\\
  \hat{\mathcal{A}}_{\ell}(\bm{\lambda})&\eqdef
  i\bigl(e^{\frac{\gamma}{2}p}\sqrt{\hat{V}_{\ell}^*(x;\bm{\lambda})}
  -e^{-\frac{\gamma}{2}p}\sqrt{\hat{V}_{\ell}(x;\bm{\lambda})}\,\bigr),\n
  \hat{\mathcal{A}}_{\ell}(\bm{\lambda})^{\dagger}&\eqdef
  -i\bigl(\sqrt{\hat{V}_{\ell}(x;\bm{\lambda})}\,e^{\frac{\gamma}{2}p}
  -\sqrt{\hat{V}_{\ell}^*(x;\bm{\lambda})}\,e^{-\frac{\gamma}{2}p}\bigr).
\end{align}
It is illuminating to compare these potential functions
\eqref{intV}--\eqref{intVs} with those of the original \eqref{Vform}
and deformed \eqref{Vl}--\eqref{Vl*} systems.
Again it is obvious that the overall normalisation of the deforming
polynomial $\xi_{\ell}$ is immaterial for the deformation.
For this choice of $\hat{\mathcal{A}}_{\ell}(\bm{\lambda})$ and
$\hat{\mathcal{A}}_{\ell}(\bm{\lambda})^{\dagger}$, one of the pair of
Hamiltonians $\hat{\mathcal{H}}_{\ell}^{(+)}(\bm{\lambda})$
\eqref{H+-} becomes proportional to the original Hamiltonian
$\mathcal{H}(\bm{\lambda})$ \eqref{origham}
with $\bm{\lambda}\to\bm{\lambda}+\ell\bm{\delta}+\bm{\tilde{\delta}}$
and the partner Hamiltonian $\hat{\mathcal{H}}_{\ell}^{(-)}(\bm{\lambda})$
is proportional to the deformed Hamiltonian
$\mathcal{H}_{\ell}(\bm{\lambda})$ \eqref{deformham}:
\begin{align}
  \hat{\mathcal{H}}_{\ell}^{(+)}(\bm{\lambda})
  &=\hat{\kappa}_{\ell}(\bm{\lambda})
  \bigl(\mathcal{H}(\bm{\lambda}+\ell\bm{\delta}+\bm{\tilde{\delta}})
  +\hat{f}_{\ell,0}(\bm{\lambda})\hat{b}_{\ell,0}(\bm{\lambda})\bigr),
  \label{Hl+=H}\\
  \hat{\mathcal{H}}_{\ell}^{(-)}(\bm{\lambda})
  &=\hat{\kappa}_{\ell}(\bm{\lambda})
  \bigl(\mathcal{H}_{\ell}(\bm{\lambda})
  +\hat{f}_{\ell,0}(\bm{\lambda})\hat{b}_{\ell,0}(\bm{\lambda})\bigr),
  \label{Hl-=Hl}
\end{align}
where $\hat{\kappa}_{\ell}(\bm{\lambda})$ and $\bm{\tilde{\delta}}$
are given by
\begin{equation}
  \hat{\kappa}_{\ell}(\bm{\lambda})\eqdef\left\{
  \begin{array}{ll}
  1&:\text{cH,\,W}\\
  (a_1a_2q^{\ell})^{-1}&:\text{AW}
  \end{array}\right.\!\!,
  \quad
  \bm{\tilde{\delta}}\eqdef\left\{
  \begin{array}{ll}
  (\frac12,-\frac12)&:\text{cH}\\[2pt]
  (\frac12,\frac12,-\frac12,-\frac12)&:\text{W,\,AW}
  \end{array}\right.\!\!.
  \label{hatkappa}
\end{equation}
The multiplicative and additive constants are common to the pair of
Hamiltonians \eqref{Hl+=H}--\eqref{Hl-=Hl}.
These are the main results of this paper.
Like the corresponding formulas in ordinary quantum mechanics \cite{stz},
these fundamental results can be obtained by explicit calculation.
The three formulas in \S\,\ref{sec:xil_prop} are essential. That is, the
difference equation satisfied by the deforming polynomial
$\xi_{\ell}(\eta(x);\bm{\lambda})$, \eqref{xildiffeq} and the two
identities relating the deforming polynomial
$\xi_{\ell}(\eta(x);\bm{\lambda})$ to its shifted one
$\xi_{\ell}(\eta(x);\bm{\lambda}+\bm{\delta})$, \eqref{xil(l+d)} and
\eqref{xil(l)}.

It is instructive to verify that the zero modes of
$\hat{\mathcal{A}}_{\ell}(\bm{\lambda})$ and
$\hat{\mathcal{A}}_{\ell}(\bm{\lambda})^{\dagger}$ do not belong to
the Hilbert space of the eigenfunctions.
In fact, the zero mode $\hat{\mathcal{A}}_{\ell}(\bm{\lambda})$ is
\begin{equation}
  \hat{\mathcal{A}}_{\ell}(\bm{\lambda})\chi=0,\quad
  \chi=\xi_{\ell}(\eta(x);\bm{\lambda})
  \phi_0(x;\mathfrak{t}(\bm{\lambda}+(\ell-1)\bm{\delta}).
\end{equation}
It has at least one pole in the rectangular domain
$x_1\leq\text{Re}\,x\leq x_2$, $|\text{Im}\,x|\leq\frac12|\gamma|$,
therefore it cannot belong to the Hilbert space.
The zero mode of $\hat{\mathcal{A}}_{\ell}(\bm{\lambda})^{\dagger}$ is
\begin{equation}
  \hat{\mathcal{A}}_{\ell}(\bm{\lambda})^{\dagger}\rho=0,\quad
  \rho=\frac{\phi_0(x;\mathfrak{t}(\bm{\lambda}+(\ell-1)\bm{\delta})^{-1}}
  {\sqrt{\xi_{\ell}(\eta(x-i\tfrac{\gamma}{2});\bm{\lambda})
  \xi_{\ell}(\eta(x+i\tfrac{\gamma}{2});\bm{\lambda})
  V^*(x-i\tfrac{\gamma}{2};\mathfrak{t}(\bm{\lambda}+(\ell-1)\bm{\delta})}}\,,
\end{equation}
which is obviously non-square integrable.
This situation is the discrete analogue of the `broken susy' case in
ordinary quantum mechanics in the terminology of supersymmetric
quantum mechanics \cite{susyqm,junkroy}. 

Based on the results \eqref{Hl+=H}--\eqref{Hl-=Hl}, we have
\begin{gather}
  \hat{\phi}_{\ell,n}^{(+)}(x;\bm{\lambda})
  =\phi_n(x;\bm{\lambda}+\ell\bm{\delta}+\bm{\tilde{\delta}}),\quad
  \hat{\phi}_{\ell,n}^{(-)}(x;\bm{\lambda})
  =\phi_{\ell,n}(x;\bm{\lambda}),
  \label{phi+-=..}\\
  \hat{\mathcal{E}}_{\ell,n}^{(\pm)}(\bm{\lambda})
  =\hat{\kappa}_{\ell}(\bm{\lambda})\bigl(
  \mathcal{E}_n(\bm{\lambda}+\ell\bm{\delta}+\bm{\tilde{\delta}})
  +\hat{f}_{\ell,0}(\bm{\lambda})\hat{b}_{\ell,0}(\bm{\lambda})\bigr)
  =\hat{\kappa}_{\ell}(\bm{\lambda})\bigl(
  \mathcal{E}_{\ell,n}(\bm{\lambda})
  +\hat{f}_{\ell,0}(\bm{\lambda})\hat{b}_{\ell,0}(\bm{\lambda})\bigr).
  \label{E+-=..}
\end{gather}
Then it is trivial to verify $\hat{\mathcal{A}}_{\ell}(\bm{\lambda})
\hat{\phi}_{\ell,n}^{(+)}(x;\bm{\lambda})\neq 0$ and
$\hat{\mathcal{A}}_{\ell}(\bm{\lambda})^{\dagger}
\hat{\phi}_{\ell,n}^{(-)}(x;\bm{\lambda})\neq 0$.
For, if one of the eigenfunction is annihilated by
$\hat{\mathcal{A}}_{\ell}(\bm{\lambda})$
($\hat{\mathcal{A}}_{\ell}(\bm{\lambda})^\dagger$), the left hand side of
\eqref{Hl+=H}(\eqref{Hl-=Hl}) vanishes, whereas the right hand side is
$\hat{\kappa}_{\ell}(\bm{\lambda})\bigl(
\mathcal{E}_n(\bm{\lambda}+\ell\bm{\delta}+\bm{\tilde{\delta}})+
\hat{f}_{\ell,0}(\bm{\lambda})\hat{b}_{\ell,0}(\bm{\lambda})\bigr)$
times the eigenfunction, which is obviously non-vanishing.
Note that $\mathcal{E}_n(\bm{\lambda}+\ell\bm{\delta}+\bm{\tilde{\delta}})
=\mathcal{E}_n(\bm{\lambda}+\ell\bm{\delta})$.

The correspondence of the pair of eigenfunctions
$\hat{\phi}_{\ell,n}^{(\pm)}(x)$ is expressed as
\begin{equation}
  \hat{\phi}_{\ell,n}^{(-)}(x;\bm{\lambda})
  =\frac{\hat{\mathcal{A}}_{\ell}(\bm{\lambda})
  \hat{\phi}_{\ell,n}^{(+)}(x;\bm{\lambda})}
  {\sqrt{\hat{\kappa}_{\ell}(\bm{\lambda})}\,\hat{f}_{\ell,n}(\bm{\lambda})},
  \quad
  \hat{\phi}_{\ell,n}^{(+)}(x;\bm{\lambda})
  =\frac{\hat{\mathcal{A}}_{\ell}(\bm{\lambda})^{\dagger}
  \hat{\phi}_{\ell,n}^{(-)}(x;\bm{\lambda})}
  {\sqrt{\hat{\kappa}_{\ell}(\bm{\lambda})}\,\hat{b}_{\ell,n}(\bm{\lambda})}.
  \label{phi+<->phi-}
\end{equation}
We introduce operators $\hat{\mathcal{F}}_{\ell}(\bm{\lambda})$ and
$\hat{\mathcal{B}}_{\ell}(\bm{\lambda})$ defined by
\begin{align}
  \hat{\mathcal{F}}_{\ell}(\bm{\lambda})&\eqdef
  \psi_{\ell}(x;\bm{\lambda})^{-1}\circ
  \frac{\hat{\mathcal{A}}_{\ell}(\bm{\lambda})}
  {\sqrt{\hat{\kappa}_{\ell}(\bm{\lambda})}}\circ
  \phi_0(x;\bm{\lambda}+\ell\bm{\delta}+\bm{\tilde{\delta}}),
  \label{hatFldef}\\
  \hat{\mathcal{B}}_{\ell}(\bm{\lambda})&\eqdef
  \phi_0(x;\bm{\lambda}+\ell\bm{\delta}+\bm{\tilde{\delta}})^{-1}\circ
  \frac{\hat{\mathcal{A}}_{\ell}(\bm{\lambda})^{\dagger}}
  {\sqrt{\hat{\kappa}_{\ell}(\bm{\lambda})}}\circ
  \psi_{\ell}(x;\bm{\lambda}).
\end{align}
The operators $\hat{\mathcal{F}}_{\ell}(\bm{\lambda})$ and
$\hat{\mathcal{B}}_{\ell}(\bm{\lambda})$ are expressed explicitly by
using the concrete forms of $V(x;\bm{\lambda})$,
$ \psi_{\ell}(x;\bm{\lambda})$ and $\phi_0(x;\bm{\lambda})$:
\begin{align}
  \!\hat{\mathcal{F}}_{\ell}(\bm{\lambda})&=
  \frac{-i}{\varphi(x)}\Bigl(
  v_1(x;\bm{\lambda}+\ell\bm{\delta})
  \xi_{\ell}(\eta(x+i\tfrac{\gamma}{2});\bm{\lambda})e^{\frac{\gamma}{2}p}
  -v_1^*(x;\bm{\lambda}+\ell\bm{\delta})
  \xi_{\ell}(\eta(x-i\tfrac{\gamma}{2});\bm{\lambda})e^{-\frac{\gamma}{2}p}
  \Bigr),\!\!\!
  \label{hatFlform}\\
  \!\hat{\mathcal{B}}_{\ell}(\bm{\lambda})&=
  \frac{1}{\xi_{\ell}(\eta(x);\bm{\lambda})}\frac{-i}{\varphi(x)}\Bigl(
  v_2(x;\bm{\lambda}+(\ell-1)\bm{\delta})e^{\frac{\gamma}{2}p}
  -v_2^*(x;\bm{\lambda}+(\ell-1)\bm{\delta})e^{-\frac{\gamma}{2}p}\Bigr).
  \!\!\!
 \label{hatBlform}
\end{align}
Here is a technical remark on \eqref{hatFlform}.
In deriving the explicit form of the operator
$\hat{\mathcal{F}}_{\ell}(\bm{\lambda})$ in \eqref{hatFlform},
one extracts the factorised potential function $v_1(x)$ and $v_1^*(x)$
from the corresponding expression of the square root of the
{\em twisted potential\/}
$\sqrt{V(x;\mathfrak{t}(\bm{\lambda}+(\ell-1)\bm{\delta}))}$ and
$\sqrt{V^*(x;\mathfrak{t}(\bm{\lambda}+(\ell-1)\bm{\delta}))}$.
Thus the choice of the argument is a subtle problem, in particular,
for the cH case, in which only one factor undergoes the sign change.
In other (W and AW) cases, sign change occur in two factors and thus
the effect cancels out.
Let us consider the twisted potential of cH case
\begin{equation*}
  V\bigl(x;\mathfrak{t}(\bm{\lambda}+\ell\bm{\delta})\bigr)
  =(-a_1-\tfrac{\ell}{2}+ix)(a_2+\tfrac{\ell}{2}+ix).
\end{equation*}
For positive $x$ close to the origin ($|x|\ll1$), we choose
$-a_1-\tfrac{\ell}{2}+ix$ to have an argument close to $+\pi$.
Then its $*$-operation $-a_1-\tfrac{\ell}{2}-ix=-(a_1+\tfrac{\ell}{2}+ix)$
has an argument close to $-\pi$. This would mean
$\sqrt{-(a_1+\tfrac{\ell}{2}+ix)^2}=-i(a_1+\tfrac{\ell}{2}+ix)
=-v_1(x;\bm{\lambda}+\ell\bm{\delta})$, instead of the naively obtained
$i(a_1+\tfrac{\ell}{2}+ix)=v_1(x;\bm{\lambda}+\ell\bm{\delta})$.

The operators $\hat{\mathcal{F}}_{\ell}(\bm{\lambda})$ and
$\hat{\mathcal{B}}_{\ell}(\bm{\lambda})$ act as the forward and backward
shift operators connecting the original polynomials $P_n(\eta)$ and
the exceptional polynomials $P_{\ell,n}(\eta)$:
\begin{align}
  \hat{\mathcal{F}}_{\ell}(\bm{\lambda})
  P_n(\eta(x);\bm{\lambda}+\ell\bm{\delta}+\bm{\tilde{\delta}})
  =\hat{f}_{\ell,n}(\bm{\lambda})P_{\ell,n}(\eta(x);\bm{\lambda}),
  \label{FhatPn=Pln}\\
  \hat{\mathcal{B}}_{\ell}(\bm{\lambda})P_{\ell,n}(\eta(x);\bm{\lambda})
  =\hat{b}_{\ell,n}(\bm{\lambda})
  P_n(\eta(x);\bm{\lambda}+\ell\bm{\delta}+\bm{\tilde{\delta}}).
  \label{BhatPln=Pn}
\end{align}
The former relation \eqref{FhatPn=Pln} with the explicit form of
$\hat{\mathcal{F}}_{\ell}(\bm{\lambda})$ \eqref{hatFlform} provides the
new explicit expression \eqref{mainres} of the exceptional orthogonal
polynomials, which is one of the main results of this paper.

Other simple consequences of these relations are
\begin{equation}
  \hat{\mathcal{E}}_{\ell,n}^{(\pm)}(\bm{\lambda})
  =\hat{\kappa}_{\ell}(\bm{\lambda})\hat{f}_{\ell,n}(\bm{\lambda})
  \hat{b}_{\ell,n}(\bm{\lambda}),\quad
   \mathcal{E}_n(\bm{\lambda}+\ell\bm{\delta})
  =\hat{f}_{\ell,n}(\bm{\lambda})\hat{b}_{\ell,n}(\bm{\lambda})
  -\hat{f}_{\ell,0}(\bm{\lambda})\hat{b}_{\ell,0}(\bm{\lambda}).
  \label{Eln+-}
\end{equation}

The normalisation constant $h_{\ell,n}(\bm{\lambda})$ \eqref{hln} of
the exceptional polynomials is related to that of the original
polynomial $h_n(\bm{\lambda})$ \eqref{hn}:
\begin{equation}
  h_{\ell,n}(\bm{\lambda})
  =\frac{\hat{b}_{\ell,n}(\bm{\lambda})}{\hat{f}_{\ell,n}(\bm{\lambda})}
  h_n(\bm{\lambda}+\ell\bm{\delta}+\bm{\tilde{\delta}})
  =\frac{\hat{b}_{\ell,n}(\bm{\lambda})}{\hat{f}_{\ell,n}(\bm{\lambda})}
  \frac{\hat{f}_{0,n}(\bm{\lambda}+\ell\bm{\delta})}
  {\hat{b}_{0,n}(\bm{\lambda}+\ell\bm{\delta})}
  h_n(\bm{\lambda}+\ell\bm{\delta}).
  \label{hln2}
\end{equation}
In the second equality we have used the explicit forms of
$h_n(\bm{\lambda})$ \eqref{hn}.
Eq.\,\eqref{hln2} is shown in the following way:
\begin{align}
  &\quad\hat{\kappa}_{\ell}(\bm{\lambda})
  \hat{f}_{\ell,n}(\bm{\lambda})\hat{f}_{\ell,m}(\bm{\lambda})
  \int_{x_1}^{x_2}dx\,\phi_{\ell,n}(x;\bm{\lambda})
  \phi_{\ell,m}(x;\bm{\lambda})\n
  &\stackrel{\text{(\romannumeral1)}}{=}
  \int_{x_1}^{x_2}dx\,\hat{\mathcal{A}}_{\ell}(\bm{\lambda})
  \phi_n(x;\bm{\lambda}+\ell\bm{\delta}+\bm{\tilde{\delta}})\cdot
  \hat{\mathcal{A}}_{\ell}(\bm{\lambda})
  \phi_m(x;\bm{\lambda}+\ell\bm{\delta}+\bm{\tilde{\delta}})\n
  &\stackrel{\text{(\romannumeral2)}}{=}
  \int_{x_1}^{x_2}dx\,\hat{\mathcal{A}}_{\ell}(\bm{\lambda})^{\dagger}
  \hat{\mathcal{A}}_{\ell}(\bm{\lambda})
  \phi_n(x;\bm{\lambda}+\ell\bm{\delta}+\bm{\tilde{\delta}})\cdot
  \phi_m(x;\bm{\lambda}+\ell\bm{\delta}+\bm{\tilde{\delta}})\n
  &\stackrel{\text{(\romannumeral3)}}{=}
  \hat{\mathcal{E}}^{(+)}_{\ell,n}(\bm{\lambda})\int_{x_1}^{x_2}dx\,
  \phi_n(x;\bm{\lambda}+\ell\bm{\delta}+\bm{\tilde{\delta}})
  \phi_m(x;\bm{\lambda}+\ell\bm{\delta}+\bm{\tilde{\delta}})\n
  &\stackrel{\text{(\romannumeral4)}}{=}
  \hat{\kappa}_{\ell}(\bm{\lambda})
  \hat{f}_{\ell,n}(\bm{\lambda})\hat{b}_{\ell,n}(\bm{\lambda})
  h_n(\bm{\lambda}+\ell\bm{\delta}+\bm{\tilde{\delta}})\delta_{nm}.
  \label{intformula}
\end{align}
Here we have used
\eqref{phi+<->phi-} and \eqref{phi+-=..} in (\romannumeral1),
\eqref{H+-Scheq} and \eqref{phi+-=..} in (\romannumeral3),
\eqref{Eln+-} and \eqref{intPnPm} in (\romannumeral4).
In order to show (\romannumeral2), we need to shift the integration
contour to the imaginary direction. It is allowed if
$\hat{V}_{\ell}(x;\bm{\lambda})\phi_0(x;\bm{\lambda}
+\ell\bm{\delta}+\bm{\tilde{\delta}})^2$ has no pole
in the rectangular domain $x_1\leq\text{Re}\,x\leq x_2$,
$0\leq\frac{\text{Im}\,x}{\gamma}\leq\frac12$.
This condition is fulfilled if the deforming polynomial
$\xi_{\ell}(\eta(x);\bm{\lambda})$ has no zero in the rectangular domain
$x_1\leq\text{Re}\,x\leq x_2$, $|\text{Im}\,x|\leq\frac12|\gamma|$,
which is indeed the case.

\subsection{Other intertwining relations}
\label{sec:other}

It is interesting to note that the operator
$\hat{\mathcal{A}}_{\ell}(\bm{\lambda})$ intertwines those of the original
and deformed systems $\mathcal{A}(\bm{\lambda})$ and
$\mathcal{A}_{\ell}(\bm{\lambda})$:
\begin{align}
  &\hat{\mathcal{A}}_{\ell}(\bm{\lambda}+\bm{\delta})
  \mathcal{A}(\bm{\lambda}+\ell\bm{\delta}+\bm{\tilde{\delta}})
  =\mathcal{A}_{\ell}(\bm{\lambda})
  \hat{\mathcal{A}}_{\ell}(\bm{\lambda}),
  \label{AhA=AlAh}\\
  &\hat{\mathcal{A}}_{\ell}(\bm{\lambda})
  \mathcal{A}(\bm{\lambda}+\ell\bm{\delta}+\bm{\tilde{\delta}})^{\dagger}
  =\mathcal{A}_{\ell}(\bm{\lambda})^{\dagger}
  \hat{\mathcal{A}}_{\ell}(\bm{\lambda}+\bm{\delta}).
  \label{AhAd=AldAh}
\end{align}
In terms of the definitions of the forward shift operators
$\mathcal{F}(\bm{\lambda})$ \eqref{Fdef},
$\mathcal{F}_{\ell}(\bm{\lambda})$ \eqref{Fldef},
$\hat{\mathcal{F}}_{\ell}(\bm{\lambda})$ \eqref{hatFldef}, and
$\mathcal{B}(\bm{\lambda})$ \eqref{Bdef},
$\mathcal{B}_{\ell}(\bm{\lambda})$ \eqref{Bldef},
the above relations are rewritten as:
\begin{align}
  &\sqrt{\hat{\kappa}_{\ell}(\bm{\lambda}+\bm{\delta})}\,
  \hat{\mathcal{F}}_{\ell}(\bm{\lambda}+\bm{\delta})
  \mathcal{F}(\bm{\lambda}+\ell\bm{\delta}+\bm{\tilde{\delta}})
  =\sqrt{\hat{\kappa}_{\ell}(\bm{\lambda})}\,
  \mathcal{F}_{\ell}(\bm{\lambda})
  \hat{\mathcal{F}}_{\ell}(\bm{\lambda}),
  \label{FlhF=FlFlh}\\
  &\sqrt{\hat{\kappa}_{\ell}(\bm{\lambda})}\,
  \hat{\mathcal{F}}_{\ell}(\bm{\lambda})
  \mathcal{B}(\bm{\lambda}+\ell\bm{\delta}+\bm{\tilde{\delta}})
  =\sqrt{\hat{\kappa}_{\ell}(\bm{\lambda}+\bm{\delta})}\,
  \mathcal{B}_{\ell}(\bm{\lambda})
  \hat{\mathcal{F}}_{\ell}(\bm{\lambda}+\bm{\delta}).
  \label{FlhB=BlFlh}
\end{align}
These relations can be proven by explicit calculation with the help of
the three formulas of the deforming polynomial
$\xi_{\ell}(\eta;\bm{\lambda})$ \eqref{xildiffeq}--\eqref{xil(l)} in
\S\,\ref{sec:xil_prop}.
 
By applying $\hat{\mathcal{A}}_{\ell}(\bm{\lambda}+\bm{\delta})$ to
\eqref{Aphi=fphi} and $\hat{\mathcal{A}}_{\ell}(\bm{\lambda})$ to
\eqref{Adphi=bphi} together with the use of \eqref{AhA=AlAh},
\eqref{AhAd=AldAh} and \eqref{phi+<->phi-}, we obtain
\begin{align}
  \mathcal{A}_{\ell}(\bm{\lambda})\phi_{\ell,n}(x;\bm{\lambda})
  &=\sqrt{\frac{\hat{\kappa}_{\ell}(\bm{\lambda}+\bm{\delta})}
  {\hat{\kappa}_{\ell}(\bm{\lambda})}}\,
  f_n(\bm{\lambda}+\ell\bm{\delta}+\bm{\tilde{\delta}})
  \frac{\hat{f}_{\ell,n-1}(\bm{\lambda}+\bm{\delta})}
  {\hat{f}_{\ell,n}(\bm{\lambda})}\,
  \phi_{\ell,n-1}(x;\bm{\lambda}+\bm{\delta})\n
  &=f_n(\bm{\lambda}+\ell\bm{\delta})
  \phi_{\ell,n-1}(x;\bm{\lambda}+\bm{\delta}),
  \label{Alphiln=fnphiln2}\\
  \mathcal{A}_{\ell}(\bm{\lambda})^{\dagger}
  \phi_{\ell,n-1}(x;\bm{\lambda}+\bm{\delta})
  &=\sqrt{\frac{\hat{\kappa}_{\ell}(\bm{\lambda})}
  {\hat{\kappa}_{\ell}(\bm{\lambda}+\bm{\delta})}}\,
  b_{n-1}(\bm{\lambda}+\ell\bm{\delta}+\bm{\tilde{\delta}})
  \frac{\hat{f}_{\ell,n}(\bm{\lambda})}
  {\hat{f}_{\ell,n-1}(\bm{\lambda}+\bm{\delta})}\,
  \phi_{\ell,n}(x;\bm{\lambda})\n
  &=b_{n-1}(\bm{\lambda}+\ell\bm{\delta})
  \phi_{\ell,n}(x;\bm{\lambda}+\bm{\delta}).
  \label{Aldphiln=bnphiln2}
\end{align}
In the calculation use is made of the explicit forms of
$\hat{\kappa}_{\ell}(\bm{\lambda})$,
$\hat{f}_{\ell,n}(\bm{\lambda})$, $f_n(\bm{\lambda})$ and
$b_n(\bm{\lambda})$ in the second equalities.
This provides a proof of \eqref{Alphiln=flnphiln}--\eqref{fln,bln}
without recourse to the shape invariance.
Likewise the above intertwining relations of the forward-backward
shift operators \eqref{FlhF=FlFlh}--\eqref{FlhB=BlFlh} give the simple
proof of \eqref{FlPln=flnPln}--\eqref{BlPln=blnPln}, respectively,
again without recourse to the shape invariance.

\section{Summary and Comments}
\setcounter{equation}{0}

The Darboux-Crum transformations intertwining the Hamiltonians of the
continuous Hahn, Wilson and Askey-Wilson polynomials with those of the
corresponding exceptional polynomials are constructed in a unified fashion.
This gives a much simpler expressions \eqref{mainres} of the exceptional
Wilson and Askey-Wilson polynomials than those given in a previous paper
\cite{os17}. The exceptional continuous Hahn polynomials are new.
See the recent work of Gomez-Ullate et al \cite{gomez2}. It provides the
iso-spectral Darboux-Crum transformations intertwining the Hamiltonians
of the $X_{\ell}$ Laguerre polynomials to that of the radial oscillator
and shows the shape invariance.

The present paper is a discrete version of the recent work \cite{stz}
which provides
the Darboux-Crum transformations intertwining the Hamiltonians of the
radial oscillator/Darboux-P\"oshl-Teller potential (the Laguerre and
Jacobi polynomials) with those of the corresponding exceptional polynomials
\cite{os16,os19,os18}. This offers a simple proof of the shape invariance
of the $\ell$-th exceptional polynomials through the established shape
invariance of the original polynomials as depicted in the following
commutative diagram:

\begin{align*}
\begin{CD}
  \fbox{\begin{tabular}{@{}c@{}}
  $\hat{\mathcal{H}}_{\ell}^{(+)}(\bm{\lambda}+\bm{\delta})$\\
  $\propto\mathcal{H}(\bm{\lambda}+(\ell+1)\bm{\delta}+\bm{\tilde{\delta}})
  +c(\bm{\lambda}+\bm{\delta})$
  \end{tabular}}
  @>{\textstyle\ \hat{\mathcal{A}}_{\ell}(x;\bm{\lambda}+\bm{\delta})\ }>>
  \fbox{\begin{tabular}{@{}c@{}}
  $\hat{\mathcal{H}}_{\ell}^{(-)}(\bm{\lambda}+\bm{\delta})$\\
  $\propto\mathcal{H}_{\ell}(\bm{\lambda}+\bm{\delta})
  +c(\bm{\lambda}+\bm{\delta})$
  \end{tabular}}\\[4pt]
  @A\begin{tabular}{@{}c@{}}\text{established}\\\text{shape invariance}
  \end{tabular}AA
  @AA\begin{tabular}{@{}c@{}}\text{shape}\\\text{invariance}
  \end{tabular}A\\[4pt]
  \fbox{\begin{tabular}{@{}c@{}}
  $\hat{\mathcal{H}}_{\ell}^{(+)}(\bm{\lambda})$\\
  $\propto\mathcal{H}(\bm{\lambda}+\ell\bm{\delta}+\bm{\tilde{\delta}})
  +c(\bm{\lambda})$
  \end{tabular}}
  @>{\textstyle{\qquad \hat{\mathcal{A}}_{\ell}(x;\bm{\lambda})\qquad}}>>
  \fbox{\begin{tabular}{@{}c@{}}
  $\hat{\mathcal{H}}_{\ell}^{(-)}(\bm{\lambda})$\\
  $\propto\mathcal{H}_{\ell}(\bm{\lambda})+c(\bm{\lambda})$
  \end{tabular}}\\[4pt]
  \text{original polynomial}
  @.
  \text{exceptional polynomial}
\end{CD}\\[6pt]
  \text{Two ways of proving shape invariance of the exceptional
  polynomials system.}\hspace{5mm}
\end{align*}

The Darboux-Crum transformation also supplies the annihilation/creation
operators for the exceptional polynomial systems through those
$a^{(\pm)}(\bm{\lambda})$ for the original system \eqref{anncreori}
\begin{equation}
  \hat{\mathcal{A}}_{\ell}(x;\bm{\lambda})
  a^{(\pm)}(\bm{\lambda}+\ell\bm{\delta}+\bm{\tilde\delta})
  \hat{\mathcal{A}}_{\ell}(x;\bm{\lambda})^\dagger,
\end{equation}
which map $\phi_{\ell,n}(x;\bm{\lambda})$ to
$\phi_{\ell,n\pm1}(x;\bm{\lambda})$.
The analogous formulas work for the exceptional Laguerre and Jacobi
polynomials \cite{stz}. As shown in \cite{os7,os13}, the
annihilation/creation operators together with the Hamiltonian constitute
dynamical symmetry algebra of an exactly solvable system, for example,
the $q$-oscillator algebra \cite{os11}. It would be an interesting
challenge to clarify the structure of the dynamical symmetry algebras
associated with the exceptional Askey type polynomials.
Likewise the three term recurrence relations for the original polynomials
are mapped to those of the exceptional polynomials \cite{hos}.
However, their significance and utility are as yet unclear.

In \cite{os18}, the shape invariance of the infinitely many potentials
for the exceptional Laguerre and Jacobi polynomials are attributed to as
many cubic identities among the original polynomials.
The corresponding identities for the cH, W and AW polynomials are
given by \eqref{shapeinvV}. These identities are quartic in $\xi_{\ell}$
and can be proven by using \eqref{xil(l+d)} and \eqref{xil(l)}.

Rodrigues type formulas for the exceptional polynomials $P_{\ell,n}$
are obtained by multiple applications of the backward shift operators for
the exceptional polynomials $\mathcal{B}_{\ell}(\bm{\lambda})$
\eqref{BlPln=blnPln} or multiple applications of the  backward shift
operators for the original polynomials $\mathcal{B}(\bm{\lambda})$
\eqref{BP=bP} followed by the intertwining forward shift operator
$\hat{\mathcal{F}}_{\ell}(\bm{\lambda})$ \eqref{FhatPn=Pln}
\begin{align}
  P_{\ell,n}(\eta(x);\bm{\lambda})
  &=\prod_{k=0}^{n-1}\frac{\mathcal{B}_{\ell}(\bm{\lambda}+k\bm{\delta})}
  {b_{n-1-k}(\bm{\lambda}+(\ell+k)\bm{\delta})}\cdot
  \xi_{\ell}(\eta(x);\bm{\lambda}+(n+1)\bm{\delta})\\
  &=\frac{\hat{\mathcal{F}}_{\ell}(\bm{\lambda})}
  {\hat{f}_{\ell,n}(\bm{\lambda})}
  \prod_{k=0}^{n-1}
  \frac{\mathcal{B}(\bm{\lambda}+(\ell+k)\bm{\delta}+\bm{\tilde{\delta}})}
  {b_{n-1-k}(\bm{\lambda}+(\ell+k)\bm{\delta})}\cdot 1,
\end{align}
where $\prod_{k=0}^{n-1}a_k=a_0a_1\cdots a_{n-1}$ is the ordered product
notation of operators.
In this connection, it is interesting to compare the forward-backward
shift operators for the original polynomials \eqref{Fdef}--\eqref{Bdef},
the exceptional polynomials \eqref{Fldef}--\eqref{Bldef} and the
intertwining ones \eqref{hatFlform}--\eqref{hatBlform}.
For the original polynomials, the forward shift operator
$\mathcal{F}(\bm{\lambda})$ \eqref{Fdef} is trivial, whereas the
potential dependence is contained in the backward shift operator
$\mathcal{B}(\bm{\lambda})$ \eqref{Bdef}.
For the exceptional polynomials, the forward shift operator
$\mathcal{F}_{\ell}(\bm{\lambda})$ \eqref{Fldef} is solely determined
by the deforming polynomial $\xi_{\ell}$ and the potential function enters
in the backward shift operator $\mathcal{B}_{\ell}(\bm{\lambda})$
\eqref{Bldef}. The intertwining ones are really twisted.
The twisted part $v_1(x)$ of the potential function enters in
$\hat{\mathcal{F}}_{\ell}(\bm{\lambda})$ \eqref{hatFlform}, whereas
the untwisted part $v_2(x)$ remains in
$\hat{\mathcal{B}}_{\ell}(\bm{\lambda})$ \eqref{hatBlform}.

Certain generating functions of the exceptional polynomials are
easily constructed from those of the original polynomials
through the main result \eqref{mainres}.
Suppose a generating function of the original orthogonal polynomials
$P_n$ is given by
\begin{equation}
  G(t,x;\bm{\lambda})=\sum_{n=0}^{\infty}\alpha_n(\bm{\lambda})
  P_n(\eta(x);\bm{\lambda})t^n,
\end{equation}
in which $\alpha_n(\bm{\lambda})$ is a constant.
For explicit forms, see \cite{koeswart}, eqs.\,(1.4.11)--(1.4.13) for cH,
(1.1.12)--(1.1.15) for W and (3.1.13)--(3.1.15) for AW.
Then eq.\,\eqref{mainres} gives the corresponding generating function
of the exceptional orthogonal polynomials $P_{\ell,n}$,
\begin{align}
  &\quad\sum_{n=0}^{\infty}
  \alpha_n(\bm{\lambda}+\ell\bm{\delta}+\bm{\tilde{\delta}})
  \hat{f}_{\ell,n}(\bm{\lambda})P_{\ell,n}(\eta(x);\bm{\lambda})t^n\n
  &=\frac{-i}{\varphi(x)}\Bigl(
  v_1(x;\bm{\lambda}+\ell\bm{\delta})
  \xi_{\ell}(\eta(x+i\tfrac{\gamma}{2});\bm{\lambda})
  G(t,x-i\tfrac{\gamma}{2};\bm{\lambda}+\ell\bm{\delta}+\bm{\tilde{\delta}})\n
  &\phantom{=\frac{-i}{\varphi(x)}}
  -v_1^*(x;\bm{\lambda}+\ell\bm{\delta})
  \xi_{\ell}(\eta(x-i\tfrac{\gamma}{2});\bm{\lambda})
  G(t,x+i\tfrac{\gamma}{2};\bm{\lambda}+\ell\bm{\delta}+\bm{\tilde{\delta}})
  \Bigr).
\end{align}

It is well known that the Laguerre polynomials are obtained from the
Jacobi polynomials in a certain limit. Likewise the Wilson polynomials
are produced by a certain limit from the Askey-Wilson polynomials.
The corresponding limiting relations for the exceptional Laguerre and
Wilson polynomials are discussed in \cite{os19} and \cite{hos}, respectively.
Here we comment on the limiting relations among the exceptional
continuous Hahn polynomials and the exceptional Wilson polynomials.
The continuous Hahn polynomials are obtained from the Wilson polynomials
in the following limit:
\begin{equation}
  \lim_{L\to\infty}(-2L)^{-n}W_n\bigl((x+L)^2;
  \alpha_1-iL,\alpha_3+iL,\alpha_2-iL,\alpha_4+iL\bigr)
  =n!\,p_n(x;\alpha_1,\alpha_2,\alpha_3,\alpha_4).
\end{equation}
By taking $x^{\text{W}}=x^{\text{cH}}+L$ and
$\bm{\lambda}^{\text{W}}=(a_1^{\text{cH}}-iL,a_1^{\text{cH}}+iL,
a_2^{\text{cH}}-iL,a_2^{\text{cH}\,*}+iL)$, and after appropriate
overall rescaling, various quantities of the exceptional Wilson systems
reduce to those of the exceptional continuous Hahn system in this
$L\to\infty$ limit.

The idea of the infinitely many exceptional Laguerre and Jacobi polynomials
\cite{os16} was obtained while studying various possibility of generating
exactly solvable quantum mechanical systems from the known shape invariant
ones, {\em e.g.} the radial oscillator and DPT potentials.
Adler's modification \cite{adler,dubov} of Crum's theorem is the most
comprehensive way to generate infinite variety of exactly solvable systems
from a known ones \cite{bagsam,junkroy}.
After the formulation of the discrete quantum mechanics version of Crum's
theorem \cite{yermzhed,matveev,os15}, its modification \`a la Adler is now
published \cite{gos}. Its Appendix has many formulas reminiscent of those
given in this paper.

There are two types of discrete quantum mechanics. In one of them, as
discussed in this paper, difference operators cause shifts in the pure
imaginary direction \cite{os13}. The formulation of the other type of
discrete quantum mechanics, in which difference operators cause real
shifts, is provided in \cite{os12}. The corresponding eigenfunctions of
the known solvable systems consist of the orthogonal polynomials of a
discrete variable, for example, the ($q$-) Racah polynomials,
\cite{koeswart,nikiforov}. It is a good challenge to construct the
exceptional polynomials corresponding to these orthogonal polynomials
of a discrete variable.

\section*{Acknowledgements}

R.\,S. is supported in part by Grant-in-Aid for Scientific Research
from the Ministry of Education, Culture, Sports, Science and Technology
(MEXT), No.19540179.


\end{document}